# Geometric gravitational origin of neutrino oscillations and mass-energy

Gustavo R. González-Martín

Departamento de Física, Universidad Simón Bolívar,
Apartado 89000, Caracas 1080-A, Venezuela.
Webpage: http://prof.usb.ve/ggonzalm

Abstract

A mass-energy scale for neutrinos was calculated from the null cone curvature using geometric concepts. The scale is variable depending on the gravitational potential and the trajectory inclination $\iota$ with respect to the field direction. The proposed neutrino covariant equation provides the adequate curvature. The mass-energy at the Earth surface varies from a horizontal value $^{h}E_0 = 0.402$ eV to a vertical value $^{r}E_0 = 0.569$ eV.

Earth spinor waves with winding numbers $n$ show differences $E_{nm}^2$ within ranges $2.05 \times 10^{-3}$ eV$^2 \leq E_{10}^2(\iota) \leq 4.10 \times 10^{-3}$ eV$^2$ and $3.89 \times 10^{-5}$ eV$^2 \leq \Delta E_{21}^2(\iota) \leq 7.79 \times 10^{-5}$ eV$^2$. These waves interfere and the different phase velocities produce neutrino-like oscillations.

The experimental results for atmospheric and nuclear neutrino oscillation mass parameters respectively fall within the $E_{10}$ and $E_{21}$ theoretical ranges. Neutrinos in outer space, where interactions may be neglected, appear as particles travelling with zero mass-energy on null geodesics. These gravitational curvature energies are consistent with neutrino oscillations, zero neutrino rest masses, Einstein's General Relativity and energy mass equivalence principle. When analyzing or averaging experimental neutrino mass-energy results of different experiments on the Earth it is of interest to consider the possible influence of the trajectory inclination angle.

04.50+h, 14.70Fm, 14.70Hp, 02.20.Qs

# 1- Introduction

Some geometric ideas about space-time may shed light on one important present-day problem in particle physics: the neutrino mass. It has been known for a long time that light deflection experiments confirm that the null cone acquires curvature under gravitation as predicted by Einstein's General Relativity. The curvature of geodesics affect particle trajectories as mass does. It may be interpreted using the energy-mass equivalence principle that particles acquire a curvature mass-energy determined by the square root of the gravitational potential absolute value. This means that weak gravitational effects should not be neglected "a priori" when studying very fast particles moving on trajectories on the curved null cone or nearby mass hyperboloids. In this article we consider the motion of a neutrino wave by discussing its differences with a light wave instead of simply assuming geodesic motion on a flat null cone. Geometric concepts and calculations are presented in appendices.

Rather than only rely on the metric structure of space-time, Einstein[1] and Schrödinger[2] maintained that it is necessary to start from fundamental affine connections in order to describe the interactions experienced by matter as it follows geometric equations of motion. The connection transformations form a group with subgroups and should unify different interactions. A modern formulation of these ideas is by a local physical action of a generalized connection[3,4] or potential $A$ as indicated in appendix A. The connection determines the straightest line between two points. If this line is also the shortest distance between the points the connection is called Riemannian. The group of the connection may be determined by the space-time Clifford algebra which acts over a generalized spin space. The connection determines a general curvature which is not necessarily Riemannian. When the interaction reduces to a gravitational field the geometric group reduces to SL(2,C), designated as L, which is homomorphic to SO(3,1) and the connection is pseudo-Riemannian.

In general the equation of motion of matter on space-time is a generalized covariant Dirac equation. Under reduction to the SL(2,C) group in vacuum this equation reduces to a covariant Weyl equation for a zero bare-mass left-handed fundamental spinor. The spinor evolution or motion may still be described as motion under a connection as E. Cartan envisioned[5]. This may be accomplished by a spinor connection valued on the sl(2,C) subalgebra and its corresponding curvature.

In his book E. Cartan proves a theorem on the impossibility of maintaining the geometric interpretation of spinors when using the original relativistic arbitrary system of curved coordinates. We should use the Cartan orthonormal moving frame formulation to display the fundamental importance of spinor geometry on relativity.

# 2- The effects of Curvature

We shall use a simple example to illustrate the influence of curvature on physical motion. Consider the motion of light along a null geodesic on the Earth surface. For the weak Schwarzchild gravitational field we may neglect the squared potential terms. The metric interval for a radial null path may be approximately written in terms of the gravitational potential $\varphi$ which determines a dimensionless Lorentzian-section curvature $K$ as discussed in appendix D,

$$d\tau^2 = (1+2\varphi)dt^2 - \frac{dr^2}{(1+2\varphi)} - r^2\left(d\theta^2 + d\phi^2 \sin^2\theta\right) = 0 ,$$

$$d\tau^2 \approx (1+2\varphi)dt^2 - (1-2\varphi)dr^2 - r^2\left(d\theta^2 + d\phi^2 \sin^2\theta\right) = 0 ,$$

$$dt^2 - dr^2 \approx -2\varphi\left(dt^2 + dr^2\right) = K\left(dr^+\right)^2 ,$$

where $r^+ = \frac{1}{\sqrt{2}}(t+r)$ is the affine parameter along the null-geodesic. This equation may also be written in a form similar to the energy-momentum relation for a particle with a squared mass proportional to the Lorentzian-section curvature

$$\left(\frac{dt}{dr^+}\right)^2 - \left(\frac{dr}{dr^+}\right)^2 = \left(\sqrt{-2\varphi}\right)^2 = K. \tag{1}$$

If the trajectory has a horizontal component the equation would be different due to the anisotropy of the Schwarzchild solution. The gravitational field determines a curvature scalar on a null Lorentzian 2-surface in a pseudo-Riemannian space which varies with position and direction of motion as indicated in appendix D. In similarity with curves on spheres the trajectories on a curved 2-surface in the curved null cone are characterized by their line curvatures which are determined by this scalar and may be related to mass-energy.

Consider the spacelike wave vector $k$ of a zero-mass particle (see fig. 1). There is a 2-dimensional null flat space spanned by the timelike $t$ and the spacelike $s$ unit vectors. This plane is the $t, s$ boost space, which contains the trajectory four-velocity $u$ and is tangent to a Lorentzian null surface. There is a well defined curvature tensor on this surface. The Gaussian curvature may be generalized on manifolds provided with moving frames and special affine transformation groups on surfaces[6,7]. Physically a Lorentzian curvature form represents the action of spacetime kinetic transformations instead of rotations. We let the physical problem determine the significant curvature involved.

On a Lorentzian surface the scalar product is not geometrically significant because it is zero for any pair of null tangent vectors. Instead we should use the geometrically significant fact that Lorentzian surfaces are characterized by pairs of null directions parametrized by the time $t$ or a space coordinate $s$ along the trajectory. We relate the curvature form to the null directions which are determined by $t^2$ equal $s^2$. Points on a null cone are characterized by points on spheres of radius $t$ related by Lorentz dilation transformations. A null direction is characterized by a point on the unit sphere $S^2$. The unit sphere $S^2$ is a cross section of the null cone[8] at the unit radius and corresponds to the celestial sphere. The observation of wave propagation is the projection of this structure on the geometry of the observer. This projection should provide a significant null-section curvature scalar.

The value of the dimensionless null-section curvature scalar determined by the projection of a photon path on an Earth trajectory in space-time is calculated in appendix D. For convenience we characterize this value by a superscript indicating the particle spin. The curvature scalar depends on the dimensionless weak gravitational potential and increases with the trajectory inclination angle relative to the horizontal plane orthogonal to the radial (vertical) direction.

## 3- An even sl(4,R) Spinorial Neutrino Equation.

There is a fundamental relation of 2-spinors with the space-time formulation of relativity, which is a remarkable geometric correspondence showing the fundamental spinor structure of light propagation. The light null-cone vectors may be constructed from squared spinors. In fact any single 2-spinor defines a real null light-like vector by the quadratic relation

$$\lambda^\mu = \overline{\zeta}^{\dot{X}} \sigma^\mu_{\dot{X}A} \zeta^A .$$

Consider now the equation of motion for a spinor. The spinor motion may be described as a particular case of the covariant Dirac equation of motion[3]. This is accomplished by a connection valued on the sl(4,R) algebra and its corresponding curvature. The equations decouple for the even and odd 2-spinor parts $(\eta,\xi)$ of the Dirac 4-spinor, when the mass $m$ is zero and the only field is the gravitational connection $\Gamma$, as in the case of flat space. The resultant equations correspond to covariant parallel transportation of the spinor fields. An even sl(4,R) or sl(2,C) generator determines a complex structure on spacetime as indicated in appendix C. We should distinguish the geometric spinor structure from its position complex coordinates on a surface in spacetime.

It is convenient to use the spacetime complex coordinate adapted to the spinor motion designated by $w$ in C and the other $z$ coordinate. The direction of motion, a null 4-velocity vector $u$, should be given in relation to the orthonormal set $\kappa^\mu$ of the Clifford algebra in order to define the null Lorentzian subspace implicit in the spinor motion. The derivatives with respect to the two complex coordinates $w$ and $z$ should also be related to the orthonormal set of the Clifford algebra. To preserve both the spinor and complex structures we should project the complex coordinate derivatives onto an orthonormal set adapted to the direction. In order to do this, we should define a set in the $R_{3,1}$ Clifford algebra[9] adapted to null vectors and complex coordinates in terms of the orthonormal set,

$$\kappa^z = \tfrac{1}{\sqrt{2}}\left(\kappa^x + i\kappa^y\right) \qquad \kappa^{\bar{z}} = \tfrac{1}{\sqrt{2}}\left(\kappa^x - i\kappa^y\right),$$

$$\kappa^w = \tfrac{1}{\sqrt{2}}\left(\kappa^s + i\kappa^l\right) \sim \tfrac{1}{\sqrt{2}}\left(\kappa^s + \kappa^t\right) \qquad \kappa^{\bar{w}} = \tfrac{1}{\sqrt{2}}\left(\kappa^s - i\kappa^l\right) \sim \tfrac{1}{\sqrt{2}}\left(\kappa^s - \kappa^t\right).$$

The desired scalar neutrino spinor equation is the projection of the spinor derivative on the spinor wave vector or equivalently on the complex tangent $w$ plane represented by $\kappa^w$. This projection determines a coordinate invariant scalar operator. Using capital letters for the complex coordinates and $\hbar=1$ physical units the neutrino equation of motion is

$$\kappa^A \nabla_A \nu = 0 \ . \tag{2}$$

Since there is no motion along the complex $z$ direction the equation of motion may be simply written

$$\kappa^A \nabla_A \nu = \kappa^w \nabla_w \nu + \kappa^{\bar{w}} \nabla_{\bar{w}} \nu = 0 \ .$$

If we take the second derivative of eq. (2) there appear quadratic operators of the form

$$\kappa^A \kappa^B \nabla_A \nabla_B \nu = 0.$$

The $\kappa^A$ anticommuting properties and the curvature symmetries determine that there are no contributions from mixed components. These terms introduce the even 4 by 4 sl(4,R) Pauli matrix $\kappa^0 \kappa^\mu$ representation of the Pauli $\sigma^\mu$ matrices. Decomposing the second derivative in its symmetric and antisymmetric parts we obtain using the commutation relations

$$\tfrac{1}{2}\{\kappa^A,\kappa^{\bar{B}}\} \nabla_A \nabla_{\bar{B}} \nu + \tfrac{1}{2}[\kappa^A,\kappa^{\bar{B}}] \nabla_A \nabla_{\bar{B}} \nu = \partial_w \partial_{\bar{w}} \nu + \kappa^0 \kappa^s {}_s\hat{\Omega}_{w\bar{w}} \nu = 0, \tag{3}$$

where the complex curvature form $\hat{\Omega}$ is determined by the commutator of the connection form. We let this physical equation determine the significant curvature for the neutrino motion.

## 3.1. A Lorentzian section curvature scalar.

In the Schwarzschild geometry the sl(2,C) complex curvature form is defined by the well known Schwarzschild curvature tensor or sl(3,1) real form.[10] The form components are equal, but the groups are homomorphic. The vector so(3,1) generator and the real volume element are replaced by the spinor sl(2,C) generator and the complex volume element,[11] as discussed in appendices B and D, according to their 1/2 homomorphic structures. Contraction of the curvature form in equation (3) with the physical components of the null vector determines a single Lorentzian null-section curvature scalar $\hat{K}$ or $\tfrac{1}{2}K$ for a 2-spinor. The curvature corresponding to a spinor on a null path along $s$ may be split into radial and horizontal components which are given, in terms of the Newtonian potential by the expressions

$$_r\Omega = \frac{2\varphi E_0^1 dr \wedge dt}{r^2} \rightarrow {}_r\hat{\Omega} \equiv \frac{2\varphi}{r^2}\left(\frac{\sigma^1}{2}\right)\frac{{}_r\hat{\Sigma}}{2} = \frac{\varphi\sigma^1{}_r(d\bar{w}\wedge dw)}{2r^2},$$

$$_h\Omega = \frac{\varphi E_0^3 d\phi \wedge dt}{r^2} \rightarrow {}_h\hat{\Omega} \equiv \frac{\varphi}{r^2}\left(\frac{\sigma^3}{2}\right)\frac{{}_h\hat{\Sigma}}{2} = \frac{\varphi\sigma^3{}_h(d\bar{w}\wedge dw)}{4r^2}.$$

The sign is chosen so the gravitational boost is attractive, along the negative radial direction. We should project these forms on the wave vector $k$ using the adapted coordinates $w$ as indicated in the curvature term of eq. (3).

Consider the laplacian differential wave operator in eq. (3). It should also be expressed in rotational coordinates as the curvature forms. For a horizontal $k$ direction the spacelike component of the operator corresponds to the angular momentum operator $L$, which without loss of generality, may be chosen along any great circle, parametrized by the angular coordinate,

$$\frac{1}{r^2}\left(\frac{\partial}{\partial r}\left(r^2\frac{\partial}{\partial r}\right)+\frac{1}{\sin\theta}\frac{\partial}{\partial\theta}\left(\sin\theta\frac{\partial}{\partial\theta}\right)+\frac{1}{\sin^2\theta}\frac{\partial^2}{\partial\phi^2}\right) = \frac{L^2}{r^2} \rightarrow \frac{1}{r^2}\frac{\partial^2}{\partial\phi^2}.$$

The angular momentum operator appears with an extra $r^2$ coordinate factor. The presence of a similar extra factor in the curvature denominator is a coordinate effect. The circular arc $r\phi$ is the rotation parameter $s$ of $L^2$ on the circle at $r$. The action of SL(2,C) is through boosts along horizontal and radial directions. We should expect a similar parameter for the curvature boost operator in terms of hyperbolic arcs $r\zeta$. Hyperbola branches should be symmetrically oriented around the center of gravitation. The distance from a hyperbola vertex or apex to its center is a constant $r$. It is convenient to choose parametric coordinates using the hyperbolic functions $r\cosh\zeta$ and $r\sinh\zeta$ on the $t,s$ subspace. We may set $w$ equal to a complex $rv$ along the wave direction. Using the $v$ complex coordinate the 4-dimensional complex wave operator eq. (3) becomes, in terms of a boost operator $\Pi^2$ on the null $w$ subspace and a curvature scalar $\hat{K}$,

$$\partial_w\partial_{\bar{w}}v + \sigma^s\hat{\Omega}_{w\bar{w}}v = 0 \rightarrow \frac{\partial_v\partial_{\bar{v}}v}{r^2} + \frac{\sigma^s\hat{\Omega}_{v\bar{v}}v}{r^2} = \frac{-\Pi^2 v}{r^2} + \frac{{}_s\hat{K}v}{r^2} = 0.$$

We explicitly indicate the coordinate role of the $r^2$ term in the curvature by defining a Lorentzian-section curvature scalar from the components of the curvature form in the last equation.

The wave vector $k$ defines a polar angle $\theta$ with respect to the gravitational field direction which determines the Lorentzian section of interest in the null-cone subspace. We obtain a single equation for the Lorentzian-section curvature scalar for spinor motion along any null direction by combining the volume element 2-forms in the curvature. The projection on null forms requires projections on both the timelike and and the spacelike components,

$$2{}_r\hat{\Sigma} + {}_h\hat{\Sigma} = -4dr\wedge dt - 2dh\wedge dt = -(4ds\wedge dt)\cos^2\theta - 2(ds\wedge dt)\sin^2\theta = (2\cos^2\theta + \sin^2\theta){}_s\hat{\Sigma},$$

$$\hat{K} \equiv \sigma^s{}_s\hat{\Omega}(\partial_{\bar{v}},\partial_v) = \frac{\varphi}{4}\left(2{}_r\hat{\Sigma} + {}_h\hat{\Sigma}\right)(\partial_{\bar{v}},\partial_v) = \frac{\varphi}{4}\left(2\cos^2\theta + \sin^2\theta\right){}_s\hat{\Sigma}(\partial_{\bar{v}},\partial_v).$$

This definition of the Lorentzian-section curvature scalar is consistent with a different definition given in appendix D, in similarity with the Gaussian curvature[12] definition on a Riemannian section. Both curvature scalars are proportional and we may write

$$\hat{K} = \frac{-\varphi}{4}\left(2\cos^2\theta + \sin^2\theta\right) = \frac{-\varphi}{4}\left(1 + \sin^2\iota\right) . \tag{4}$$

The Laplace-Beltrami operator on the $w\bar{w}$ complex surface in the complex $T\hat{M}$ bundle,

$$\Pi^2 v = {}_s\hat{K}(r)v , \tag{5a}$$

is consistent with a laplacian on a null surface in the *TM* bundle.

The curvature scalar varies according to eq. (4) from a maximum at the vertical direction to a minimum at the horizontal direction. These results are also obtained using the matrices of the $R_{1,3}$ Clifford algebra[9] or Dirac's matrices.

Using mass ratios relative to the fundamental geometric energy-mass $\mathfrak{m}$, indicated in appendix A, the operator in eq. (5) may be mapped and interpreted as a Klein-Gordon energy-momentum operator $P^2$ with the Planck constant $\hbar$ equal to unity,

$$\Box v \to P^2 v = \frac{1}{4}\left(\left(\frac{\partial}{\partial t}\right)^2 - \left(\frac{\partial}{\partial s}\right)^2\right)v = {}_s m^2 v . \tag{5b}$$

If we restore the Planck constant in the equation, the neutrino energy operator is

$$^\nu E = \tfrac{1}{2} i\hbar \partial_t = \frac{h}{V(U(1))} i\partial_t .$$

Planck's energy relation is satisfied for the natural 2-spinor frequency which is half the vector frequency because the spinor linear group SU(2) volume is twice the volume of the orthogonal group SO(3). This may explain the difference with eq. (1) for the photon where the relevant Lorentzian curvature scalar for the vector (spin 1) is $^1K$ instead of a different $^{½}K$ for a 2-spinor.

## 4- Neutrino Mass-energy.

In a previous work discussed in appendix F it was proposed that the leptons correspond to topological excitation[13] waves characterized by a winding (or rather wrapping) topological number which may determine leptonic flavor levels. We calculated the lepton bare mass ratios[14,15] in terms of algebraic volume ratios of the group symmetric subspaces. Similarly we may define characteristic potential energy scale factors for neutrino trajectory curvature ratios. An energy factor also determines the line curvature corresponding to the trajectory of the spin-1 vector representation or photon on the null-space. The energy factor may be calculated from the dimensionless curvature scalars.

The different relative potential energy factors for each representation trajectory determine proportional line curvatures. For the spin representation the curvature scalar is given by the SL(2,C) connection instead of the homomorphic SO(3,1) connection as indicated in appendix D.1. The dimensionless line curvatures which determine spinor wave parameters are given by eq. (4),

$$^{½}K = \hat{K} = \frac{-\varphi}{2} = \frac{^1K}{4} ,$$

$$^{½}\kappa = \sqrt{\frac{-\varphi}{2}} = \frac{\sqrt{gR}}{\sqrt{2}c} = \frac{\sqrt{9.8066\,m/s^2 \times 6.3781 \times 10^6\,m}}{\sqrt{2} \times 2.9979 \times 10^8\,m/s} = 1.319 \times 10^{-5} . \tag{6}$$

The line curvatures define a potential mass-energy scale for neutrinos. In the previous lepton work we calculated the lepton mass ratios using the fundamental geometric inverse length or mass $\mathcal{M}$ indicated in appendix A. We now use the neutrino electron mass ratio given in appendix F to determine the spinor mass-energy value for a radial (vertical) motion from its radial line curvature in terms of the dimensionless gravitational potential,

$$\frac{m_\nu}{m_e} = \frac{\tfrac{1}{2}\kappa\,\mathcal{M}}{C_R\,\mathcal{M}} = \frac{\tfrac{1}{2}\kappa}{C_R}\;.$$

We obtain the energy factors from the line curvature ratios using the volumes of the same previously used symmetric spaces related to the spinor bundle. In particular the geometric factor $C_R$, indicated in appendix F, is the factor of proportionality of the electron bare rest mass with respect to a geometric mass-energy $\mathcal{M}$. This factor gives the physical value of $\mathcal{M}$ in terms of the electron mass,

$$^rE_0 = \tfrac{1}{2}\kappa\mathcal{M} = \frac{m_e\sqrt{-\varphi}}{\sqrt{2}V(C_R)} = \frac{3 m_e\sqrt{-\varphi}}{16\pi\sqrt{2}}\;.$$

On the Earth surface this effect is higher than expected because it depends on the square root of the dimensionless gravitational potential. The numerical value on the Earth surface may be taken as a constant of the order of the expected electron neutrino mass,

$$^hE_0^2 \le (1+\sin^2\iota)\frac{-3^2\,\varphi m_e^2}{2^{10}\pi^2} \le {}^rE_0^2,\tag{7}$$

$$^rE_0 = \frac{m_e\sqrt{-\varphi}}{\sqrt{2}V(C_R)} = \frac{5.11\times 10^5 \times 2.638\times 10^{-5}}{16\sqrt{2}\pi/3} = 0.569 \text{ eV} \sim m_\nu\;.\tag{8}$$

The different horizontal curvature scalar is a half of the radial curvature scalar because the Schwarzschild metric curvature scalar is not isotropic (see appendix D). It gives the minimum value of mass-energy for a horizontal neutrino trajectory on the Earth surface,

$$^hE_0 = \frac{^rE_0}{\sqrt{2}} \equiv E_0 = \frac{m_e\sqrt{-\varphi}}{2V(C_R)} = 0.402 \text{ eV} \sim m_\nu\;.\tag{9}$$

We take the characteristic mass-energy scale equal to this minimum value. Mass-energy increases with the trajectory inclination angle relative to a horizontal plane.

If gravitation is neglected there may be significant errors in the calculation of small rest masses at very-high-energy processes. The errors would be of the order of the energy related to the linear curvature as indicated in appendix E. Some of these theoretical curvature effects were first reported[16] within the context of neutrino velocity determinations.

## 5- Squared Energy Effects due to Winding Numbers.

The group of the connection also produces a gravitational action on neutrinos of different type. The same lepton model[13] used in the previous section represents the muon and tauon neutrinos as topological excitations of the electron neutrino. This implies variations of the neutrino equation determining variations of the geometric mass-energy range and the wavefunctions for these excited neutrinos.

In this section we shall restrict the discussion to spinors. Therefore we drop the spinor/vector superscript notation. The action of the geometric potential $A$ (sl(2,C) connection) on the covariant Weyl wave equation is similar to the action of rotation generators on gravitational precession[17] experiments.

The SL(2,C) group acts on its spin representations. In particular we are considering representations induced by its SU(2) subgroup, designated as H, which are spinor functions on the SL(2,C)/SU(2) coset. This 3-dimensional coset space is related to the null cone. The SL(2,C) neutrino representations have zero bare rest mass and are geometrically related to the massive leptons[18,19] through the homotopy groups of SL(2,C), Sp(4,R) and SU(2) as indicated in appendix F.

The third homotopy group of a space is defined as the classes of mappings from the 3-spheres to the space. These homotopy groups are isomorphic to the integers Z. The number $n$ of produced images is called the winding or wrapping number. The images in H allow alternative actions of SU(2) and additional neutrino wave equations on the images which we call channels $n = 0, 1, 2$.

The number of energy-degenerate group elements in each wound space level is the H group volume $V$. The squared mass-energy, which is proportional to the curvature scalar $K$, should be a characteristic parameter of any neutrino wave channel. Since all group elements are equivalent under the H group they equally share the curvature energy at any given point and direction on space-time. As a model we may consider that there are neutrino energy-mass currents along channels or parallel paths in space-time. Each parallel path corresponds to a neutrino of different type.

The volume of the spin subgroup V(H) represents a geometric inertial opposition to the energy flow. This is due to the H group action, directly on the L/H coset, which corresponds to an action of the group H on its inverse $H^{-1}$. These actions, the trajectories and energy scalars for waves vary discretely from their $n = 0$ values. The variation when other channels are present must be due to the varied number of group elements under the action of the Z homotopy group.

The variation of the volume in the neutrino family, together with the presence of non-zero curvature, produces a variation of the neutrino energy flow per channel state. We may take $K$ and the volume $V_0$ as constants but there are volume variations for the other channels. The available potential energy for all states in each of the $n$ active channels is constant and determines that the energy varies inversely proportional to the volume as a function of $n$ when $n$ varies discretely from zero,

$$K \mathfrak{M}^2 V_0^2 = E_0^2 V_0^2 = E^2(n) V^2(n).$$

The SU(2) group H acts through the connection as a derivative. We may assume that the discrete mass-energy variations produced by the relative volume variations due to the action of the homotopy group Z are small. Since the variations are small we define a map from the homotopy group Z to a discrete subset of approximate volume difference operators $\Delta$ in the set of variation operators with a derivative $\partial$ as operation,

$$Z = (1, 2, 3 \cdots n) \to \left( \Delta_V^n \equiv \frac{V^2 \partial^n (1/V^2)}{\partial V^n} \right),$$

$$\left( \Delta_V^1, \Delta_V^2 \cdots \Delta_V^n \right) = \left( -2V^{-1}, (-1)^2 6V^{-2}, \cdots (-1)^n (n+1)! V^{-n} \right).$$

The resultant $\Delta_V^n$ difference operators determine discrete squared energy differences for waves in each channel,

$$\Delta_V^1 (E_0^2) = \frac{-2K}{V} \Rightarrow \Delta E_{10}^2 = \frac{-2E_0^2}{V},$$

$$\Delta_V^2(E_0^2) = \frac{6K}{V^2} \Rightarrow \Delta E_{21}^2 = \frac{6E_0^2}{V}.$$

A wave current in a level should imply currents in all lower levels. Due to the proportionality of the equations, the energy in the channels and their differences satisfy the relations

$$E_1^2 = E_0^2 + E_{10}^2 = E_0^2 \left(1 - \frac{2}{V(SU(2))}\right), \qquad (10)$$

$$E_2^2 = E_1^2 + E_{21}^2 = E_0^2 \left(1 - \frac{2}{V(SU(2))} + \frac{6}{V^2(SU(2))}\right) \approx E_0^2 \left(1 - \frac{2}{V(SU(2))}\right) \approx E_1^2, \qquad (11)$$

$$\Delta E_{01}^2 = E_0^2 - E_1^2 = \frac{2E_0^2}{16\pi^2} \approx \Delta E_{02}^2, \qquad (12)$$

$$\Delta E_{21}^2 = E_2^2 - E_1^2 = \frac{6E_0^2}{(16\pi^2)^2}. \qquad (13)$$

For any pair of $n$ channels the neutrino wave transitions may include terms of the form

$$T = e^{i\Delta k_\mu x^\mu} + e^{-i\Delta k_\mu x^\mu} = 2\cos(\Delta k_\mu x^\mu) = 4\sin^2(\Delta k_\mu x^\mu / 2)$$

which would indicate transition oscillations among the neutrino channels.

Observed very small experimental neutrino masses and neutrino oscillations[20] may be caused by this gravitational effect. Instead of constant neutrino masses we really would have the null cone gravitational curvature and small mass-energy differences determined by variable neutrino winding numbers under the curvature effect.

The potential neutrino energies may be found from eqs. (9, 10, 11). In particular, the neutrino "dressed" mass in experiments on the Earth surface would depend on its spacelike direction because the Schwarzschild metric null-section curvature scalar is not isotropic. We express the numerical energy results in terms of the trajectory inclination angle (see fig. 1) within their horizontal minimum and vertical maximum, using the standard Earth gravitational values which may be taken as constants. We have, respectively, for Earth neutrinos associated with the short-range energy $E_{01}^2 \approx E_{02}^2$ or the long-range energy $E_{21}^2$,

$$2.05 \times 10^{-3} \text{ eV}^2 \leq \Delta E_{01}^2(\iota) = \frac{2E_0^2(1 + \sin^2 \iota)}{16\pi^2} \leq 4.10 \times 10^{-3} \text{ eV}^2, \qquad (14)$$

$$3.89 \times 10^{-5} \text{ eV}^2 \leq \Delta E_{21}^2(\iota) = \frac{6E_0^2(1 + \sin^2 \iota)}{(16\pi^2)^2} \leq 7.79 \times 10^{-5} \text{ eV}^2. \qquad (15)$$

It may be considered that these neutrino channels with gravitational energy determined by eqs. (10, 11) are energy eigenstates with wavefunctions $\nu$. They are characterized by the winding numbers and may be taken as the neutrino mass states of the standard theory[20] of neutrino oscillations.

In general we can say that the neutrino mass-energy relations would be variable, depending on the gravitational potential in each experiment. A neutrino is created by a nuclear reaction in a definite state.

From this initial state neutrino waves are driven and completely determined by the null-cone curvature scalar until they are destroyed by a final nuclear collision. If the curvature scalar is zero, neutrino waves in all channels move equally free on a flat null cone. Neutrinos in outer space, where interactions may be neglected, would have zero line curvatures and would appear as massless particles travelling as photons on flat null geodesics.

Atmospheric neutrino experiments[20] indicate a result, 17% higher than the minimum energy value of the short-range-oscillation. An average trajectory inclination may produce a variation of the curvature scalar and sufficient to account for the difference.

The reactor and solar neutrino experiments indicate a combined result[21], 2% lower than the maximum value of the long-range-oscillation. The solar neutrino wave mass-energy has a diurnal anisotropy due to the Earth rotation. There is a variation of the curvature scalar which may sufficiently increase the observed beyond the average value.

## 6- Conclusions.

The gravitational potential determines the null-section curvature scalar on the null cone which defines fundamental squared energy differences among the topological neutrino states. It also determines the proportional line curvatures of the photon and neutrino trajectories.

We calculated the neutrino mass-energy from the line curvature using dimensionless ratios of volumes of symmetric SL(4,R) coset spaces, previously used to calculate particle bare-rest-mass ratios, and the compact subgroup SU(2) of the gravitational SL(2,C) group. The experimental results of neutrino mass-energy measurements and oscillations on trajectories appear to be consistent with these theoretical results and their physical interpretation without the need to assign bare rest masses to these particles.

In other words, gravitational curvature lens effects are consistent with null-cone neutrino oscillations. There is no clear experimental distinction between mass-energy in the neutrino motion on a curved null cone and mass in its motion on a nearby hyperboloid in flat space-time.

The neutrino curvatures and energies determined by the geometry produced by gravitation are really variables. These results are due to the presence of the gravitational potential term in the space-time Schwarzschild metric. At the Earth surface the effect on the dimensionless radius of curvature of null cone trajectories is of order $10^{-5}$. When averaging experimental mass-energy values it is necessary to consider the possibility of dependence with respect to trajectory inclination. If there are additional weak fields on space-time there may be a higher effective potential and the numerical results may increase. In general for stronger fields a unified geometry would produce stronger particle effects.

The energy-mass associated to the neutrino topological excitations is related to the gravitational curvature. Bohr might have been correct when he presented the idea that neutrinos may be related to gravitation, as indicated in appendix E. A neutrino is created by a nuclear reaction in a definite state. From this state neutrino waves are driven and consistently determined by the null cone curvature scalar until they disappear in another reaction.

We may have looked for the graviton in the wrong place, associated to the metric. The metric is given by a vector frame determined by squared spinors as discussed in appendices D and E. If we look at the space where the gravitational curvature and potential act we find the spin-½ neutrino which may be considered an excitation of a spinor frame field and therefore related to the metric. Instead of a metric wave we have a neutrino wave eq. (5) with energy proportional to the frequency $p_0 = hf$ which may represent the real graviton. This is consistent with the fact that gravitational and neutrino fields are the only ones not blocked by the Earth crust.

This work is in line with previous theoretical work. Our method is not based directly on the metric properties of space-time but rather on the affine properties of matter transformations as its excitations evolve in space-time. The generalized theory, presented over a period of years, is based on geometric ideas introduced by Einstein, Cartan and Schrödinger. Further theoretical analysis and experiments may be required to fully understand the physical effects of the spinor geometry associated to particles.

# Appendix A

Traditionally General Relativity has been expressed using differential geometry. General coordinates are conveniently used to find solutions, specially for static and stationary gravitational fields. On the other hand, mathematicians moved away from general coordinates looking for a more geometrical approach[22,23] to a modern differential geometry. In particular, Elie Cartan elaborated[24,25,26] the method of the "repere mobile" or moving frames. A frame is an ordered basis for a vector space. A moving frame is a function whose values are frames in the various tangent spaces $M_p$ at points $p$ in a manifold $M$. A moving frame is not necessarily related to a proper coordinate system since the frame vectors need not commute. Later Koszul introduced a (Koszul) connection[27] which is a function $D$ associated to vector fields which satisfies certain axioms. This is a generalization for the classical connection in a coordinate system. Finally Ehresmann[28,29] completed the modern concept of connections using the action of a Lie group on the geometric elements. The use of gauge fields in Physics is another example of the applications of these geometric ideas to generally non Riemannian spaces.

In parallel to this development, Cartan also introduced the notion of spinors[30]. This concept related to Clifford algebras was applied to quantum physics by the introduction of the Pauli and Dirac matrices and later to General Relativity and Gravitation by Penrose[31].

A unified relativistic theory of gravitation, electromagnetism and other interactions[3,4] was developed along these ideas. The geometric structure group SL(4,R) of the connection is determined by the space-time Clifford algebra which acts over a generalized spin space. The field equation is a generalized Maxwell equation with a matter current $J$ as source. When the interaction reduces to only a gravitational field, the geometric group SL(4,R) reduces in a limit to SL(2,C) which is homomorphic to SO(3,1), the connection is pseudo-Riemannian and the tensor calculus reduces to the standard Ricci calculus. The geometric and algebraic modernization was a long process and it would be inappropriate to present a treatment in this article. We should limit ourselves to present the fundamental references.

The connection is a 1-form valued in the Lie algebra. The curvature is a 2-form valued in the Lie algebra which may be expressed as

$$\Omega = E_i^j \Omega_j^i$$

in terms of a standard basis $E$ in the Lie algebra and standard 2-forms. There is a fundamental geometric inverse length or mass-energy $\mathcal{M}$ related to the generalized curvature which may be defined in terms of the connection and the current[7,12],

$$\mathcal{M} = \tfrac{1}{4}\operatorname{tr}\left(J^\mu A_\mu\right) \equiv J \bullet A \ .$$

# Appendix B.

We present here an explicit display of the rotational transformation differences between vectors and spinors using, for convenience, a language close to relativistic physics rather than Cartan's geometric language.

The Lorentz SL(2,C) connection and covariant derivative in spinor space define the SO(3,1) connection and covariant derivative in curved space-time. The morphism between the connections is determined by the homomorphism between the Lorentz SO(3,1) and SL(2,C) groups. In general, the six connection 1-forms shown in appendix D correspond to the generators which form a basis in the Lie algebra of both groups. In particular, we now display this relation using the connecting Pauli matrices,

$$l_{\hat{\nu}}^{\hat{\mu}} = \text{tr}\left(\sigma^{\hat{\mu}} g^\dagger \sigma_{\hat{\nu}} g\right) \quad l_{\hat{\nu}}^{\hat{\mu}} \in \text{SO}(3,1); \ g \in \text{SL}(2,\mathbb{C}) \ .$$

The one dimensional spinor subgroup generated by $\sigma_3$ is

$$\Lambda = \exp(\beta\sigma_3) = I\cosh\beta + \sigma_3\sinh\beta = I + \beta\sigma_3 + O(\beta^2) + \cdots$$

and the corresponding Lorentz one-dimensional subgroup which may be written as a *t-z* subspace transformation is

$$L_{\hat{\nu}}^{\hat{\mu}} = \text{tr}\left(\sigma^{\hat{\mu}} \exp(\beta\sigma_3) \sigma_{\hat{\nu}} \exp(\beta\sigma_3)\right)$$

$$L_{\hat{\nu}}^{\hat{\mu}} = \begin{bmatrix} \cosh 2\beta & 0 & 0 & \sinh 2\beta \\ 0 & 1 & 0 & 0 \\ 0 & 0 & 1 & 0 \\ \sinh 2\beta & 0 & 0 & \cosh 2\beta \end{bmatrix} = I_{\hat{\nu}}^{\hat{\mu}} + 2\beta \begin{bmatrix} 0 & 0 & 0 & 1 \\ 0 & 0 & 0 & 0 \\ 0 & 0 & 0 & 0 \\ 1 & 0 & 0 & 0 \end{bmatrix} + O(\beta^2) + \cdots \approx I_{\hat{\nu}}^{\hat{\mu}} + 2\beta E_3^0 \ .$$

It is clear that this spinor generator corresponds to the Lorentz boost generator whose action on the asymptotic light-cone zone dilates the time and distance coordinates on the cone. The amount of deformation produced by the corresponding SO(3,1) connection form along the null geodesic is twice the amount of the deformation produced by the SL(2,C) connection form on the original spinor, because it also includes the deformation of the conjugate spinor. Equal expressions hold for the three spinor generators in the non-compact boost sector of the algebra. Similar but antisymmetric expressions, with the hyperbolic functions replaced by the circular functions, hold for the three spinor generators in the compact rotation sector. These relations are not due to a pure coincidence, they are determined by the homomorphism of the related groups associated to the gravitational connection. In general the group transformations generated by the connections are related by the 2 to 1 relationship between the two indicated connection group transformations.

The SL(2,C) and SO(3,1) groups are 2 to 1 homomorphic[32]. This means that the parameter space for SL(2,C) is twice the size of the parameter space for SO(3,1). A single transformation in SO(3,1) is equivalent to a pair of transformations in SL(2,C) and actually splits in two transformations: a spinor transformation and its similar conjugate transformation. The corresponding transformation produced by the Cartan-Ehresmann connection on a single spin-1/2 representation has a relative 1/2 factor with respect to the usual spin-1 vector representation transformation. The same relation respectively applies to their compact subgroups U(1) and SO(2). A factor of 2, well known for vector and spinor rotations, is introduced by the $4\pi$ volume of U(1) which is twice the volume of SO(2) and affects the definition of spinor frequency *f*. It also is a common factor for the non-compact Lorentz group generators, in particular for those which generate the time dilations on the light cone and the boosts inside the cone.

# Appendix C

The use of differential and spinor analysis displays the fundamental principles involved in the neutrino problem. We apply the spinor moving frame formulation under the SL(4,R) structure group of the geometry as presented in the theory.[19] The local physical action is through the spinor connection.

The interactions are determined by the sl(4,R) connection generators which form a basis in the $R_{3,1}$ Clifford algebra. When the interaction reduces to only a gravitational field the group reduces to its even SL(4,R) subgroup or SL(2,C). The fundamental representation of SL(2,C) is a zero-mass 2-spinor particle which has the properties of a neutrino. Within the theory this appears as the simplest geometric structure associated to space-time.

The even sl(4,R) connection generators on a manifold $M$ include antisymmetric tensors $J$ of type (1, 1) which obey $J^2 = -1$ determining an almost complex structure on a real 4-vector space[11]. The corresponding complex vector space has 2 complex dimensions. This structure $J$ determines an integral almost complex structure on $TM$ and therefore a unique complex structure on the manifold $\widehat{M}$. The manifold with the complex structure $\widehat{M}$ and the original $M$ are related manifold representations of spacetime with points *locally* identified by the physical spacetime events. The curved complex manifold is a valid representation of spacetime capable of representing a physical relativistic local expression in terms of time and space variables related to complex 2-spinors.

There should be a local bundle map $j$, which allows a physical interpretation on $M$ and the definition of complex coordinates associated to the coordinates on $M$. The 2 to 1 mapping from 2-spinors to 4-vectors determines a *local* relation between the Lie algebras of the linear transformations on $T\widehat{M}$ and $TM$. Covariant equations under an SL(2,C) connection on the curved spinor complex bundle provide an expression for the spinor motion.

We use a Minkowskian complex time coordinate as indicated in appendix D and introduce an arbitrary spacelike direction $s$ along the neutrino motion on spacetime $M$. A complex differential form basis and its dual complex vector basis may be obtained by pulling back these coordinates. Thus we define adapted complex coordinates on the 2-dimensional curved manifold with index summation also including conjugate indices.

If we have a spinor frame field we have a section in the spinor bundle and a map given by the neutrino current

$$j: \widehat{M} \to M ,$$

$$z^1 = \operatorname{Re} z^1 + \operatorname{Im} z^1 \equiv j^* s + i(j^* l) \equiv w ,$$

$$z^2 = \operatorname{Re} z^2 + \operatorname{Im} z^2 \equiv j^* x + ij^* y \equiv z .$$

The tangent plane may be perceived as decomposed into a visual null plane $w$ and a transversal complex plane $z$. The $w$ null plane is tangent to a Lorentzian surface adapted to the neutrino null motion. The volume element on the complex[11] spinor subspace spanned by $w$ which corresponds to the boost space is, omitting the complex coordinate indices,

$$\tfrac{1}{2}\Sigma = dw \wedge d\bar{w} = (j^* ds + ij^* dl) \wedge (ds - ij^* dl) = -2i(j^* ds \wedge j^* dl) = 2(ds \wedge dt) = 2\,{}^I\Sigma .$$

This means that the volume (area) element on this spinor complex tangent subspace determines twice the volume (area) element on the corresponding 2-dimensional vector tangent space. The dual complex vector base[11,28] on this complex null subspace is

$$\frac{\partial}{\partial w} = \frac{1}{2}\left(\frac{\partial}{\partial(j^* s)} - i\frac{\partial}{\partial(j^* l)}\right),$$

$$\frac{\partial}{\partial \bar{w}} = \frac{1}{2}\left(\frac{\partial}{\partial(j^* s)} + i\frac{\partial}{\partial(j^* l)}\right).$$

The desired sl(2,C) valued spinor curvature and connection forms on a subspace corresponding to any space-time subspace may be obtained from the vector curvature and connection forms by replacing the vector generator and form with the corresponding spinor generator and form.

# Appendix D

We obtain the curvature for a weak Schwarzschild space-time in terms of the Newtonian gravitational potential using Minkowskian coordinates with the fourth coordinate $l = it$, $c = 1$,

$$d\tau^2 = -ds^2 = -g_{\mu\nu}dx^\mu dx^\nu = (1+2\varphi)dl^2 + \frac{dr^2}{(1+2\varphi)} + (dx^2)^2 + (dx^3)^2 .$$

The relevant non-zero classical connection coefficients[33] (Christoffel symbols) may be calculated from the metric. The Cartan connection 1-forms $\omega$, which correspond one-to-one to the Lorentz SO(3,1) generators are obtained from the orthonormal tetrad $\theta^{\hat{\beta}}$,

$$d\theta^{\hat{\alpha}} + \omega^{\hat{\alpha}}_{\hat{\beta}} \wedge \theta^{\hat{\beta}} = 0 ,$$

$$\theta^{\hat{0}} = \sqrt{1+2\varphi}\,dt = \sqrt{1-\frac{2GM}{c^2 r}}\,dt , \qquad \theta^{\hat{1}} = \theta^{\hat{r}} = \frac{dx^1}{\sqrt{1+2\varphi}} \approx \sqrt{1+\frac{2GM}{c^2 r}}\,dr ,$$

$$\theta^{\hat{2}} = dx^2 , \qquad \theta^{\hat{3}} = dx^3 .$$

The well known six Riemann curvature 2-forms are[10]

$$\Omega = D\omega = d\omega + \omega \wedge \omega$$

$$\Omega^0_1 = \Omega^0_r \approx \varphi'' dr \wedge dt + O(\varepsilon) \approx \frac{-2GM}{c^2 r^3} dr \wedge dt = 2\varphi \frac{dr \wedge dt}{r^2} ,$$

$$\Omega^0_2 \approx \frac{\varphi''}{2} dx^2 \wedge dt + O(\varepsilon) \approx \frac{-GM}{c^2 r^3} dx^2 \wedge dt = \varphi \frac{dx^2 \wedge dt}{r^2} ,$$

$$\Omega^0_3 \approx \frac{\varphi''}{2} dx^3 \wedge dt + O(\varepsilon) \approx \frac{-GM}{c^2 r^3} dx^3 \wedge dt = \varphi \frac{dx^3 \wedge dt}{r^2} ,$$

$$\Omega^1_2 = \Omega^r_2 \approx -\varphi'' dx^2 \wedge dr + O(\varepsilon) \approx \frac{GM}{c^2 r^3} dx^2 \wedge dr = -\varphi \frac{dx^2 \wedge dr}{r^2} ,$$

$$\Omega^2_3 \approx -\varphi'' dx^3 \wedge dx^2 + O(\varepsilon) \approx \frac{2GM}{c^2 r^3} dx^3 \wedge dx^2 = -2\varphi \frac{dx^3 \wedge dx^2}{r^2} ,$$

$$\Omega^3_1 = \Omega^3_r \approx -\varphi'' dr \wedge dx^3 + O(\varepsilon) \approx \frac{GM}{c^2 r^3} dr \wedge dx^3 = -\varphi \frac{dr \wedge dx^3}{r^2} .$$

These six forms correspond to the six SO(3,1) generators which may be denoted by the following matrices

$$-E_0^1 \equiv \begin{bmatrix} 0 & 1 & 0 & 0 \\ 1 & 0 & 0 & 0 \\ 0 & 0 & 0 & 0 \\ 0 & 0 & 0 & 0 \end{bmatrix}, \quad -E_0^2 \equiv \begin{bmatrix} 0 & 0 & 1 & 0 \\ 0 & 0 & 0 & 0 \\ 1 & 0 & 0 & 0 \\ 0 & 0 & 0 & 0 \end{bmatrix}, \quad -E_0^3 \equiv \begin{bmatrix} 0 & 0 & 0 & 1 \\ 0 & 0 & 0 & 0 \\ 0 & 0 & 0 & 0 \\ 1 & 0 & 0 & 0 \end{bmatrix},$$

$$-E_1^2 \equiv \begin{bmatrix} 0 & 0 & 0 & 0 \\ 0 & 0 & 1 & 0 \\ 0 & -1 & 0 & 0 \\ 0 & 0 & 0 & 0 \end{bmatrix}, \quad -E_2^1 \equiv \begin{bmatrix} 0 & 0 & 0 & 0 \\ 0 & 0 & 0 & 0 \\ 0 & 0 & 0 & 1 \\ 0 & 0 & -1 & 0 \end{bmatrix}, \quad -E_3^1 \equiv \begin{bmatrix} 0 & 0 & 0 & 0 \\ 0 & 0 & 0 & -1 \\ 0 & 0 & 0 & 0 \\ 0 & 1 & 0 & 0 \end{bmatrix}.$$

The Schwarzschild metric curvature is not isotropic. The curvature boost components along a radial direction or its orthogonal (horizontal) directions may be written in dimensionless coordinates by

$$\Omega_1^0 = 2\varphi E_1^0 \frac{dr \wedge dt}{r^2},$$

$$\Omega_2^0 = \varphi E_2^0 \frac{dx^2 \wedge dt}{r^2},$$

$$\Omega_3^0 = \varphi E_3^0 \frac{dx^3 \wedge dt}{r^2}.$$

The $r^2$ divisor is a coordinate effect. The null geodesic section of interest is spanned by the wave vector and the time direction (see fig.1). Vectors in the directions of the wave vector and time define the null-section curvature scalar. The curvature form should be projected and evaluated along these vectors as indicated in section 2,

$$\Omega_s^0 = -\left( \sin\iota \begin{bmatrix} 0 & 1 & 0 & 0 \\ 1 & 0 & 0 & 0 \\ 0 & 0 & 0 & 0 \\ 0 & 0 & 0 & 0 \end{bmatrix} \frac{2\varphi dr}{r^2} + \cos\iota \begin{bmatrix} 0 & 0 & 1 & 0 \\ 0 & 0 & 0 & 0 \\ 1 & 0 & 0 & 0 \\ 0 & 0 & 0 & 0 \end{bmatrix} \frac{\varphi dh}{r^2} \right) \wedge dt.$$

The dimensionless curvature scalar is expressed as

$$^1K \equiv \Omega_s^0\left(\left(s\frac{\partial}{\partial s}\right),\left(t\frac{\partial}{\partial t}\right)\right) = \Omega_s^0\left(t\left(\sin\iota\frac{\partial}{\partial r}+\cos\iota\frac{\partial}{\partial h}\right),t\frac{\partial}{\partial t}\right).$$

We should set the components equal to the unit timelike and spacelike vectors to obtain the dimensionless curvature scalar.

$$^1K = -2\varphi\sin^2\iota - \varphi\cos^2\iota = -\varphi(1+\sin^2\iota).$$

The curvature scalar increases with the trajectory inclination angle relative to a horizontal direction $h$,

$$^1K_h \leq -\varphi(1+\sin^2\iota) \leq {}^1K_r.$$

## D.1. Differences with the photon motion.

We now further discuss the differences in curvature because their role in determining mass-energy. We restricted to gravitation by restricting to the Lorentz group connection on Cartan moving frames. The tangent space-time *TM*, with its complex structure, is a curved space (it is really a curved fiber bundle). The curvature operators are base-space 2-forms valued on their respective group Lie algebras as indicated in appendix A,

$$\Omega = E_i^j \Omega_j^i .$$

Since both elements $E_i^j$, and $\Omega_j^i$, in the curvature correspond to bivectors related to rotations or 2-forms, the symmetry of the resultant curvature tensor components (between the first and last index pairs) corresponds to an interchange of the rotational roles played by them. This means that both elements are linear combinations of bivectors related to infinitesimal rotations or generators in their respective sl(2,C) or so(3,1) representations.

The infinitesimal group transformation on a spin-½ (spinor) representation has a relative ½ factor with respect to the usual spin-1 (vector) representation transformation as indicated in appendix B. Similarly the volume (area) element on the spinor complex tangent subspace determines twice the volume (area) element on the corresponding 2-dimensional vector tangent space as indicated in appendix C. Due to the quadratic dependence of the curvature on these two bivectors there are two ½ factors among the respective curvatures. A dimensionless Lorentzian-section curvature scalar *K* given on *TM* determines a spinor-section curvature scalar $\hat{K}$,

$$\frac{\hat{K}}{{}^1 K} = \frac{{}^{1/2} K}{{}^1 K} = \left(\frac{1}{2}\right)^2 = \frac{\left({}^{1/2}\kappa\right)^2}{\left({}^1\kappa\right)^2} .$$

There is a relative ½ factor between the principal line curvatures in the spinor and vector spaces. The line curvatures of the principal geodesic directions affect particle trajectories as mass. These simple relations are in accordance of more detailed geometric calculations. We find the spinor values

$$ {}_h\hat{K} \leq {}_s\hat{K} = \frac{-\varphi}{4}\left(1+\cos^2\theta\right) = \frac{-\varphi}{4}\left(1+\sin^2\iota\right) \leq {}_r\hat{K} .$$

The Lorentzian-section curvature varies with the position and the trajectory polar angle relative to the gravitational field direction.

# Appendix E

In order to discuss the motion of a neutrino in the proper physical context we should first make some theoretical-historical considerations. Cartan introduced geometric spinors[5,29] in 1913. Weyl[34] soon developed a physical theory for a 2-component (Weyl) spinor. Pauli used the spin matrices to formulate the theory of non relativistic spin[35] in 1927. Dirac developed the standard 4-component relativistic spinor theory for massive fermions in 1928. It is now well known that if the particle has zero mass the Dirac equations decouple into 2-component Weyl zero mass spinor equations[36] which may be written in terms of the Pauli matrices and the Planck constant as

$$i\hbar\bar{\sigma}^\mu\partial_\mu\eta = 0 .$$

Since the introduction of a (neutrino) particle[37] in beta decay by Pauli in 1930 the main question debated was whether its mass was zero or very small. Traditionally the Weyl equation was used to describe the neutrino. Today the bound for the electron neutrino mass is estimated to be of the order of a few eV and questions arise about the adequate equation to use. It is also well known that the experimental mass of particles is generally explained by field energy corrections added to an assumed finite "bare" particle mass to obtain a theoretical "dressed" mass in the process of mass renormalization[38,39] for Dirac's equation. If the curvature effects of the gravitational potential term were equivalent to the effects of a small finite dressed mass-energy we similarly may explain the small neutrino mass starting from a zero "bare" neutrino mass. Or equivalently, the neutrino oscillations[40,41] could also be explained by the dressed mass-energy contribution to a zero bare mass neutrino. Both points of view may be consistent with observed neutrino oscillations. Certainly these questions are still open for further experiments and theoretical ideas. From a theoretical point of view it appears that the Lee and Yang two-component theory of the neutrino[42] using the zero mass Weyl equation, together with the Einstein's energy-mass equivalence should be the fundamental ideas. When there is no gravitational field the photon curve becomes a flat space-time null geodesic. Under these conditions the neutrino, as determined by the standard Weyl equation, should also follow the same path of the photon. Nevertheless when there is a gravitational field and space is curved we expect[43] the neutrino spinor field to obey a geometric covariant equation determined from the integrability conditions of the full sl(4,R) field equation restricted to the gravitational spinor sl(2,C) connection 1-form. This would be in accordance with the old Bohr idea[44,45] that the neutrino should be related to gravitation. The general equation[46] of motion reduces to a generalized covariant Weyl equation written in terms of the orthonormal basis of the Clifford algebra,

$$\kappa^\mu \nabla_\mu \nu = 0 \,.$$

This equation determines a covariant Klein-Gordon 4-momentum equation on the spinor bundle over space-time.

There is a well established classical spinor analysis on curved spinor spaces and a curved space-time.[47,48,49,50] Using the Cartan analysis the spinor connection may be expressed as sl(2,C) Lie algebra valued 1-forms. It is known that the three Pauli matrices over the complex numbers may be considered as the six generators of the SL(2,C) group of transformations which is homomorphic to the SO(3,1) Lorentz group of transformations.

# Appendix F

Mass may be defined in an invariant manner in terms of energy[7], depending on the connection and the moving matter frames ("repere mobile") in a geometric theory[4,12]. All bare rest masses depend on a geometric mass $\mathfrak{m}$ which may be normalized to the proton mass. In particular the proton to electron bare rest mass ratio is $6\pi^5$. It has been shown that the quotient of the bare rest masses of all known leptons are equal, up to correction terms of the order of the alpha constant, to the quotient of geometric masses corresponding to subgroup excitations including algebraic and topologic effects. In general, the geometry determines the geometric excitation mass spectrum, which for low masses, essentially agrees with the physical particle mass spectrum. The masses of the leptons increase under the action of a strong connection (relativity of energy) and are related to meson and quark excitation masses. There are massive connection excitations whose masses correspond to the weak boson masses and allow a geometric interpretation of Weinberg's angle. The necessary first order corrections, due to the interaction of the excitations, are of the order of the alpha constant, equal to the order of the discrepancies.

In particular it was also proposed[6] that massive leptons are representations of the Sp(4,R) group by functions characterized by rest mass and spin over the mass shell hyperboloid subspace while neutrinos are representations of the SL(2,C) group as functions characterized by helicity over the null hypercone subspace. The masses of leptons correspond to topological excitation waves characterized by a winding (or rather wrapping) topological number *n* which may determine leptonic flavor levels. The *n*-homotopy group of a group is defined as the classes of mappings from the *n*-spheres to the group space with the operation of juxtaposition of the sphere images. The third homotopy groups of SL(2,C), Sp(4,R) and SU(2) are all[11,32] isomorphic to Z and define winding or wrapping numbers *n*.

The mass of leptons is proportional to the number of states (different momentum values) which is proportional to the volume of a total geometric space related to the bundle *S(M)*, which depends on the wrapping number *n* and may be calculated,

$$V(C^T) = V(U(1)) \times (V(C_R)^{n+1}) \quad n \neq 0,$$

$$V(C_R) = \frac{16\pi}{3}.$$

where $C_R$ is the space of relativistic inequivalent points of the coset Sp(4,R)/SL(2,C). The winding or wrapping number *n* determines the number of images. The bare masses corresponding to the topological *n*-excitations are proportional to the volumes

$$V(C^T) = 4\pi \left(\frac{16\pi}{3}\right)^{n+1} \quad 0 < n \leq 2,$$

which may be expressed in terms of the electron bare mass.

For excitations with wrappings 1 and 2 we obtain the corresponding bare mass ratios of the $\mu$ and $\tau$ leptons. The lepton masses are determined by *n* and therefore *n* also characterizes lepton massive representations.

The neutrino and electron geometric mass-energy under a gravitational field are expected to have small corrections[18,19] proportional to the field line curvature $\kappa$. The total mass-energy ratio of these two particles under gravitation would be

$$\frac{m_\nu}{m_e} = \frac{\left(V\left(\frac{SL(2,C)}{SL(2,C)}\right)_R + \kappa\right)\mathfrak{M}}{\left(V\left(\frac{S_p(4,R)}{SL(2,C)}\right)_R + \kappa\right)\mathfrak{M}} = \frac{\kappa}{(V(C_R) + \kappa)} \approx \frac{\kappa}{V(C_R)}.$$

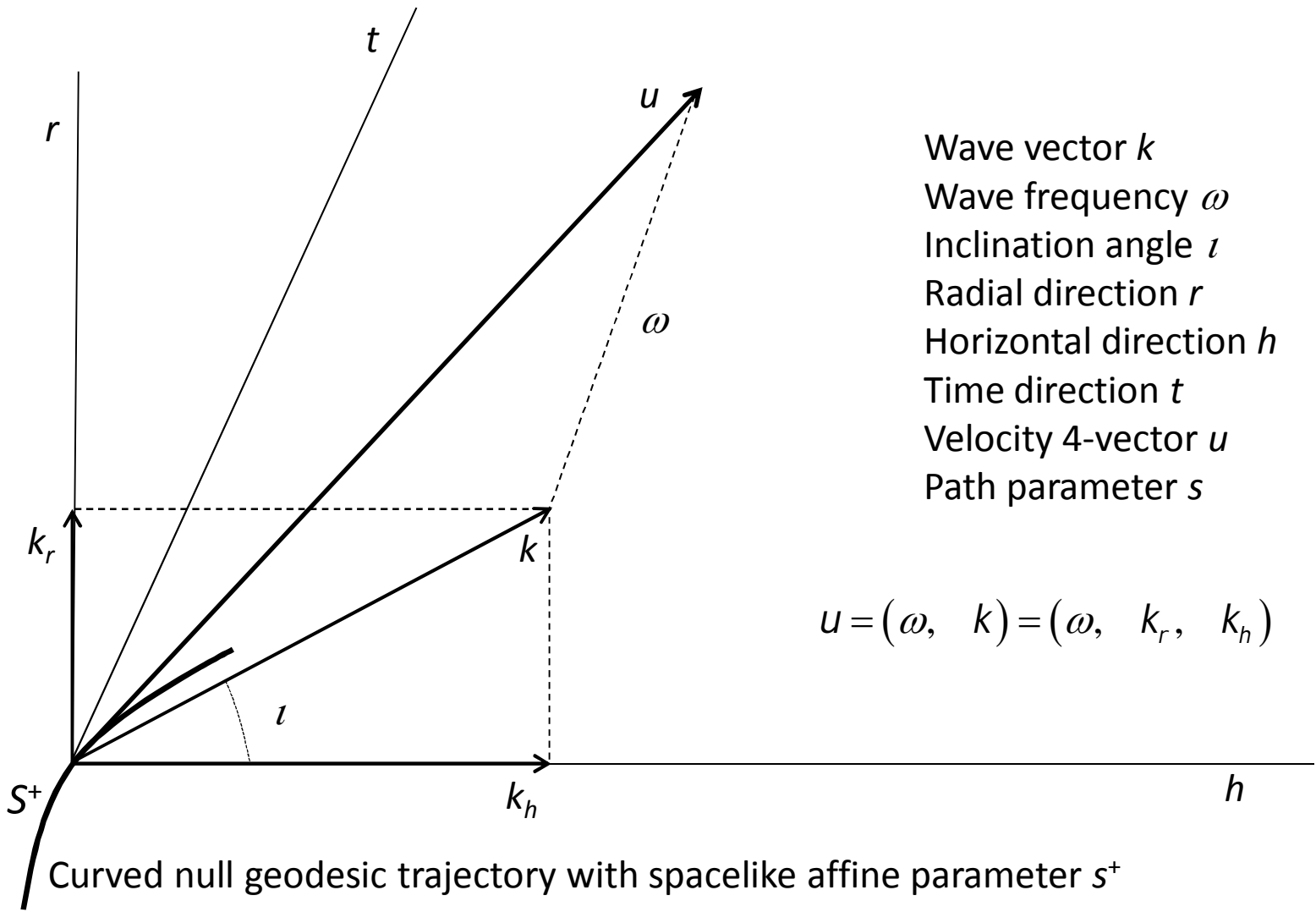

Curved null geodesic trajectory with spacelike affine parameter $s^+$

Figure 1